\documentclass{article}

\usepackage[final,nonatbib]{tackling_climate_workshop_style}

\usepackage[utf8]{inputenc} % allow utf-8 input
\usepackage[T1]{fontenc}    % use 8-bit T1 fonts
\usepackage{hyperref}       % hyperlinks
\usepackage{url}            % simple URL typesetting
\usepackage{booktabs}       % professional-quality tables
\usepackage{amsfonts}       % blackboard math symbols
\usepackage{nicefrac}       % compact symbols for 1/2, etc.
\usepackage{microtype}      % microtypography
\usepackage{wrapfig}
\usepackage[utf8]{inputenc} % allow utf-8 input
\usepackage[T1]{fontenc}    % use 8-bit T1 fonts
\usepackage{hyperref}       % hyperlinks
\usepackage{url}            % simple URL typesetting
\usepackage{booktabs}       % professional-quality tables
\usepackage{amsfonts}       % blackboard math symbols
\usepackage{nicefrac}       % compact symbols for 1/2, etc.
\usepackage{microtype}      % microtypography
\usepackage{xcolor}         % colors
\usepackage{xspace}
\usepackage{subcaption}
\usepackage{caption}
\usepackage{enumitem}
\usepackage{chemformula}
\usepackage{graphicx}
\usepackage{hyperref}
\usepackage{wrapfig}

\hypersetup{
    colorlinks=true,
    linkcolor=blue,
    filecolor=magenta,      
    urlcolor=cyan,
    pdftitle={Overleaf Example},
    pdfpagemode=FullScreen,
    }

\newcommand{\mat}[1]{\ensuremath{\mathbf{#1}}}

\newcommand{\YP}[1]{{\color{blue}[YP: #1]}}

\title{Understanding Opinions Towards Climate Change on Social Media}

\author{%
\quad Yashaswi Pupneja$^{1,2}$ \quad Joseph Zou$^3$ \quad Sacha L\'evy$^{4}$ \quad Shenyang Huang$^{1,3}$ \\
\quad $^1$Mila \quad $^2$Université de Montréal \quad $^3$McGill University \quad $^4$Yale University\\
\texttt{yashaswi.pupneja@umontreal.ca} \quad \texttt{yuesong.zou@mail.mcgill.ca}\\
\texttt{sacha.levy@yale.edu} \quad \texttt{shenyang.huang@mail.mcgill.ca} \\
}

\begin{document}

\maketitle

\begin{abstract}
    Social media platforms such as Twitter (now known as X) have revolutionized how the public engage with important societal and political topics. Recently, climate change discussions on social media became a catalyst for political polarization and the spreading of misinformation. In this work, we aim to understand how real world events influence the opinions of individuals towards climate change related topics on social media. To this end, we extracted and analyzed a dataset of 13.6 millions tweets sent by 3.6 million users from 2006 to 2019. Then, we construct a temporal graph from the user-user mentions network and utilize the Louvain community detection algorithm to analyze the changes in community structure around Conference of the Parties on Climate Change~(COP) events. Next, we also apply tools from the Natural Language Processing literature to perform sentiment analysis and topic modeling on the tweets. 
    Our work acts as a first step towards understanding the evolution of pro-climate change communities around COP events. Answering these questions helps us understand how to raise people's awareness towards climate change thus hopefully calling on more individuals to join the collaborative effort in slowing down climate change. 
\end{abstract}

\section{Introduction}

Climate change is a widely debated topic, and a significant number of individuals still denies the effect of climate change. Hence, we seek to understand how real-world events would influence opinions regarding climate change on social media platforms. According to a survey by the Pew Research Center, 15\% of adults in the US do not believe that the Earth is warming, and 36\% believe that it is due to natural causes rather than human activity. The spread of misinformation and propaganda campaigns on social media has contributed to skepticism around climate change~\cite{funk2019us}.

To address these challenges, we aim to analyze the growth or contraction of communities of climate change supporters and non-supporters on social media platforms. This analysis will provide insights into how real-world events such as the Conference of the Parties~(COP) events shape public opinion on climate change. By analyzing the patterns of social media interactions and the spread of climate change discourse on social media platforms, effective communication strategies can be developed to promote climate change awareness and counter misinformation campaigns. The stakeholders of this study include policymakers, environmental organizations, and the general public.

\section{Background and Related Work}

\textbf{Climate Change on Social Media.}
Social media have created new arenas for public debates and revolutionized the discussion of prominent issues such as global climate change~\cite{pearce2019social}. Various research on this topic have been conducted. 
\cite{effrosynidis2022exploring} looked into user profile, tweet attitude, discussion topics, and climate phenomenons. 
\cite{beiser2018assessing} built a random forest (RF) model to predict people's attitude towards climate change and environment protection based on psychological and demographic factors. Furthermore, 
\cite{tuitjer2021social} modeled the relationship between perceived climate change efficacy and news and user activities on social media by using multilevel regression. 

\textbf{Temporal Graph Learning.} 
Social networks can be modeled as a temporal graph where users are nodes and interactions are edges while the nodes and edges in the graph would naturally change over time~\cite{rossi2020temporal,zhou2022tgl,you2022roland}. The dynamic community detection task which aims to identify groups of individuals who are closely connected to each other on the temporal graph~\cite{qin2020mining,clark2019network,boudebza2020detecting, rossetti2018community}. Qin et al.~\cite{qin2020mining} and Boudebza et al.~\cite{boudebza2020detecting} both proposed novel methods to detect \emph{stable communities}, groups of nodes forming a coherent community throughout a period of time. In this work, we first apply the well-know Louvain community detection algorithm~\cite{blondel2008fast} to identify the growth or contraction of user communities of climate change topics on twitter. 

\textbf{Misinformation detection.} Misinformation related to climate change is a significant problem that has been widely researched. Several studies have examined the relationship between social media usage, belief in conspiracy theories, and misinformation, highlighting the gravity of the issue \cite{social_media_usage}. The adverse impact of such misinformation has been well-documented, underscoring the importance of inoculating people against misinformation related to climate change \cite{misinformation}. Research on debunking strategies has produced mixed results, highlighting the continued influence of misinformation and the need for effective measures~\cite{debiasing, debunking}. The identification of psychological factors that drive misinformation and hinder debunking \cite{ecker2022psychological} indicates the complexity of the issue and the need for targeted interventions to tackle it.

\textbf{Political polarization detection.} Political polarization \cite{avinash2007polictical_doi:10.1073/pnas.0702071104} is one of the crucial socio-political issues in twenty-first century. It intensifies political debates and may even threaten civil society. \cite{avinash2007polictical_doi:10.1073/pnas.0702071104} demonstrated that political polarization happens naturally when the observable indicators of policy outcomes are not monotonic. Hence due to its commonness and harmfulness, it is important to monitor political polarization and its impact especially in public discussions about climate. 
 \cite{four-party_doi:10.1080/09644016.2014.976485} demonstrated that public opinion toward climate politics is closely consistent with people's positions on parties and the US general public has become quite politically polarized in recent years. Further studies \cite{CHEN2021102348,doi:10.1080/09644016.2015.1090371} showed similar alignment between polarized public opinions toward climate politics and  partisan sorting. Considering the effect of events, 
 \cite{falkenberg2022growing} highlighted a large increase in polarization in 2021 during the 26th United Nations Conference of the Parties on Climate Change (COP26) and attributed it to growing right-wing activity. 
Moreover, \cite{9381419} investigated hostile communication  between  polarized competing groups.

\section{Dataset}
In this section, we first discuss how we constructed the twitter dataset which we used for this project in Section~\ref{sub:collect}. Then we extract the user-user mentions temporal network from the full dataset to examine the user interactions on twitter~\ref{sub:construct}. 

\subsection{Collecting Twitter Dataset} \label{sub:collect}

To understand how people engage with climate change related topics on social media, we first collect dataset of tweets from Twitter, a popular social network for discussing trending topics. In particular, we focus our collection process on climate change related tweets, posts made by users on Twitter. We leveraged a recently released \href{https://www.kaggle.com/datasets/deffro/the-climate-change-twitter-dataset}{Kaggle dataset} ~\cite{effrosynidis2022exploring, effrosynidis2022climate} containing climate change related tweets and their attributes for our study. However, the dataset lacks the raw tweet text and other information. Therefore, we use the tweet ID from the dataset to recollect the tweets using the \href{https://developer.twitter.com/en/products/twitter-api/academic-research}{Twitter API} for Academic Research. Our dataset contains 13.6 millions tweets sent by 3.6 million users from 2006 to 2019. Each tweet contains various information such as the sender, any mentions of other users, timestamp, any hashtags and the raw text of the tweet. We have also collected dates for relevant Conference of the Parties~(COP) Events in the duration of the dataset (as seen in Appendix~\ref{app:cop_events}).

\subsection{Constructing Temporal Graph} \label{sub:construct}

From the collected meta-data of tweets, we construct the user-user mentions network to be a weekly temporal graph. 
This network naturally shows how users interact with each other over time. However, the edges might not necessarily indicate that two users shares the same opinion. For example, it is possible for a climate change supporter to mention another user who doesn't believe in climate change. In this work, we discretize the mentions network to be a weekly temporal graph. More specifically, we represent the temporal graph $\mathbf{G}$ as a series of graph snapshots, $\mathbf{G} = \{ \mat{G}_t\}_{t=1}^{T}$ where $\mat{G}_t = \{ \mat{V}_t, \mat{A}_t\}$ is the graph snapshot at time $t$ and $\mat{V}_t, \mat{A}_t$ represent the set of nodes and the adjacency matrix of $\mat{G}_t$ respectively. 

While processing the tweet dataset, we find that there exist a large number of users which only send climate related tweets sparsely. Therefore, the constructed temporal graph is disconnected and sparse thus forming isolated components rather communities of users. To increase the density of the network while focusing on users that are more active on climate change topics, we remove all users that have less than 100 edges (across all steps) from the network (both outgoing and incoming edges). In this way, we only preserve a dense core set of users.
\section{Methodology}

% \begin{figure}[t]
% \centering

% \begin{subfigure}{0.5\columnwidth}
%  \includegraphics[width=\columnwidth,trim={0in 0in 0 0 },clip]{figs/hashtag_count.pdf}
%   \caption{Distribution of top Hashtags in tweets.}
%   \label{fig:hashtags}
% \end{subfigure}%
% %\hfill
% \begin{subfigure}{0.5\columnwidth}
%   \centering
%   \includegraphics[width=\columnwidth,trim={0in 0in 0 0 },clip]{figs/sentiment_analysis.png}
%   \caption{Distribution of \emph{positive}, \emph{negative} and \emph{neutral} tweets.}
%   \label{fig:sentiment_image}
% \end{subfigure}%

% \caption{Analysis of a). hashtags and b). sentiment in extracted tweets.}
% \vskip -0.2in
% \end{figure}

% \usepackage{graphicx}
% \usepackage{subcaption}

\begin{figure}[ht]
\begin{subfigure}{0.5\textwidth}
  \centering
  \includegraphics[width=\textwidth, trim={0in 0in 0in 0in}, clip]{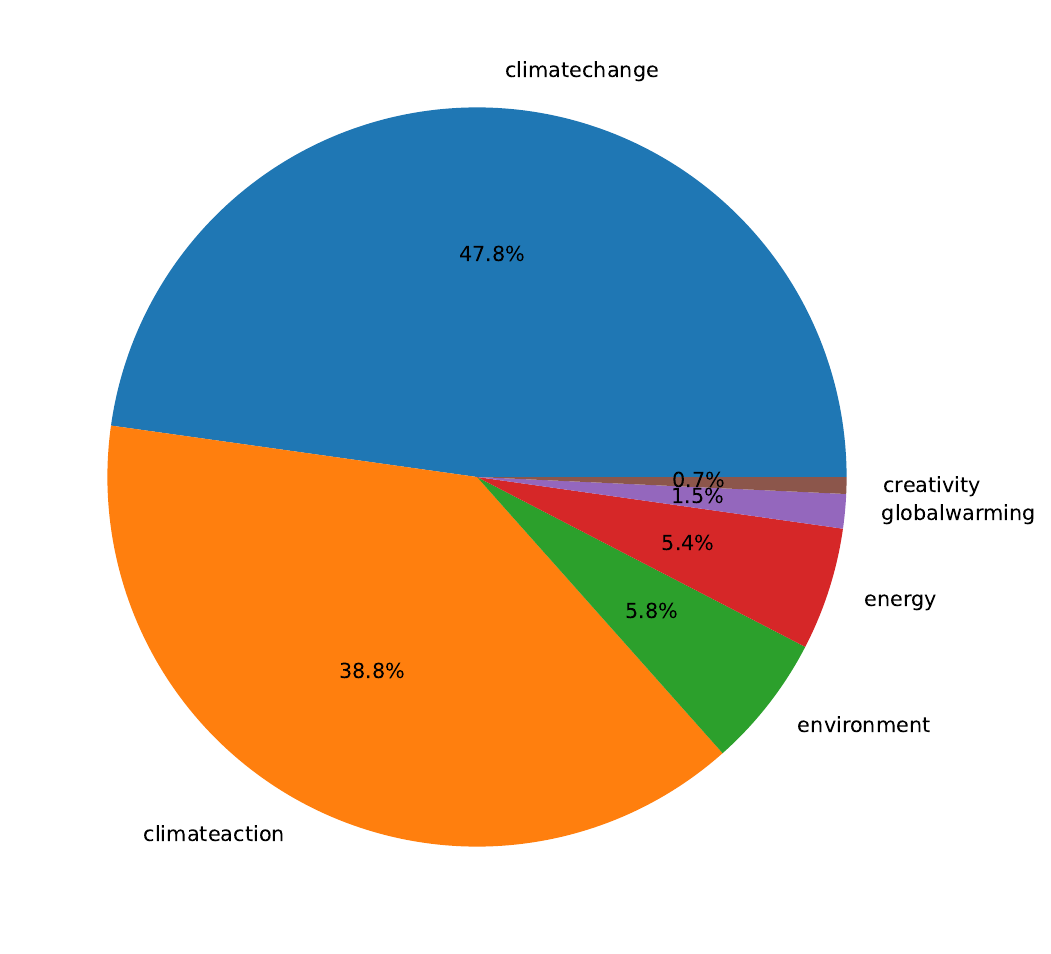}
  \caption{Distribution of top Hashtags in tweets.}
  \label{fig:hashtags}
\end{subfigure}%
\begin{subfigure}{0.5\textwidth}
  \centering
  \includegraphics[width=\columnwidth,trim={0in 0in 0 0 },clip]{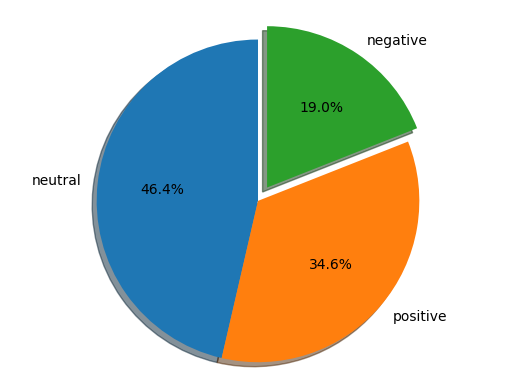}
  \caption{Distribution of \emph{positive}, \emph{negative} and \emph{neutral} tweets.}
  \label{fig:sentiment_image}
\end{subfigure}
\caption{Analysis of a). hashtags and b). sentiment in extracted tweets.}
\end{figure}

\textbf{Hashtags Analysis.} The top \YP{8} most used hashtags from tweets are shown in Figure~\ref{fig:hashtags}. Some interesting hashtags include creativity, spirituality, economy, education and "its time to change". In particular, the hashtags themselves might already contain views or opinion towards climate change topics such as "its time to change".

%Additionally, sentiment analysis is used to monitor the impact of real-world events on public opinion towards climate change. For example, sentiment analysis can track how the emotional tone of social media posts changes after a natural disaster, political decision, or climate-related news event. 

\textbf{Sentiment Analysis.}
Sentiment analysis is a technique used to determine the emotional tone behind a piece of text. We analyze the overall sentiment of tweets to understand the emotions associated with different topics related to climate change such as climate policy, environmental disasters or scientific research. Sentiment analysis also helps identify patterns in the emotional tone of different groups, such as climate change supporters or deniers.
Here, we used the \href{https://textblob.readthedocs.io/en/dev/}{TextBlob library} to detect the sentiments of the tweets which determined whether the texts were \emph{positive}, \emph{negative} or \emph{neutral}. The results are summarized in the pie chart in Figure~\ref{fig:sentiment_image}. The majority of tweets are neutral while there remains 19.0\% of tweets that have a negative sentiment. This shows that most users share a neutral sentiment towards climate change and it calls for more actions to engage the public in more positive involvement in climate change actions. We also acknowledge that negative sentiments such as fear and discontent are also considered as negative but can show awareness towards climate events. 

%we represent the temporal graph $\mathbf{G} = \{ \mat{G}_t\}_{t=1}^{T}$ as a series of graph snapshots $\mat{G}_t = \{ \mat{V}_t, \mat{A}_t\}$ where $\mat{V}_t, \mat{A}_t$ represents the set of nodes and the adjacency matrix of $\mat{G}_t$.

\textbf{Temporal Graph Analysis} 

\begin{figure}[t]
\centering
\begin{subfigure}{.4\textwidth}
  \centering
  \includegraphics[width=\linewidth]{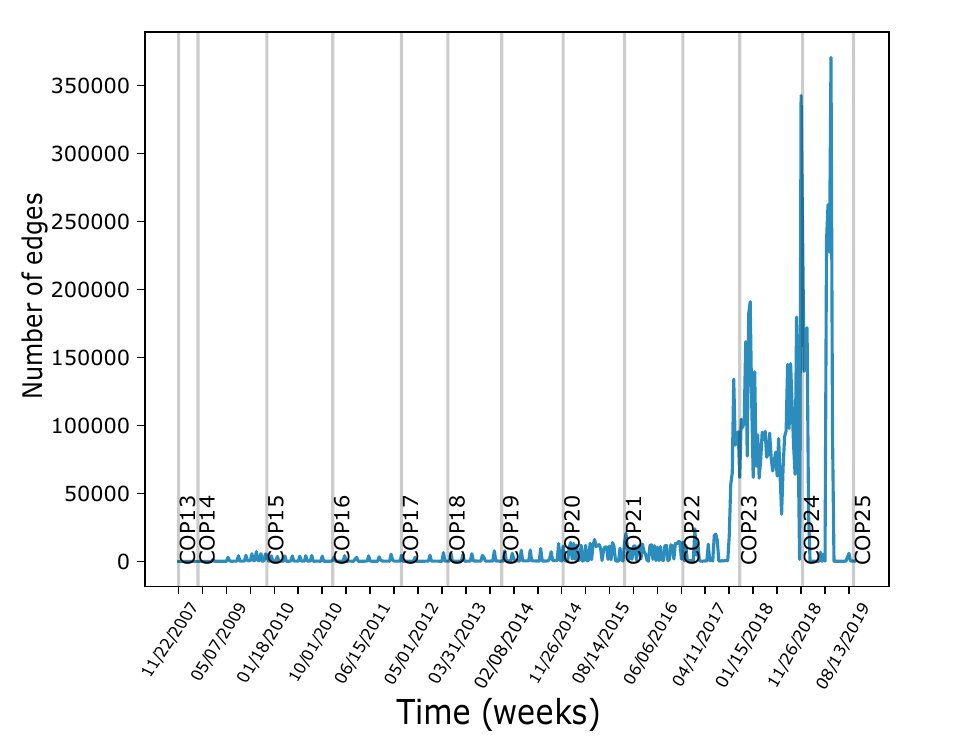}
  \caption{number of edges} 
  \label{Fig:num_edges}
\end{subfigure}
\begin{subfigure}{.4\textwidth}
  \centering
  \includegraphics[width=\linewidth]{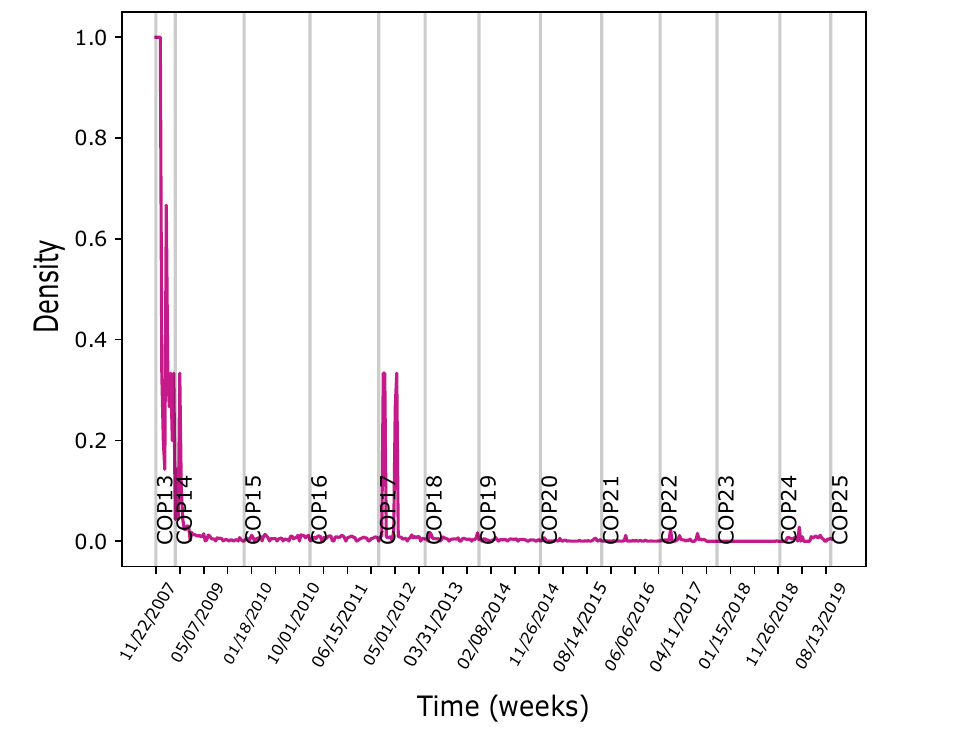}
  \caption{density of graph snapshots}
  \label{Fig:density}
\end{subfigure}%
\caption{
The evolution of a). the number of edges and b). graph density in the temporal user-user mentions network constructed from tweets.}
\label{Fig:stats}
% \vskip -0.2in
\end{figure}

Figure~\ref{Fig:stats} shows the evolution of the number of edges and graph density over time. We also marked COP events by grey vertical lines. COP events are often accompanied by local peaks in the number of edges. This is likely because more discussion on climate change related topics happen around COP events annually. For graph density, we observe large peaks around COP 13, 14 and 17. The peaks around COP 13 and 14 can be a result of noise in data collection in early years of twitter.  The peaks around COP 17 signals increased discussion between active twitter users around this time. This coincides with the fact that COP 17 has one of the most significant achievements as all countries agree to the reduction of emissions, including the US and emerging countries such as Brazil, China, India and South Africa~\cite{achieve}. Details on the time of COP events are in Appendix~\ref{app:cop_events}.

% \subsection{Community Detection on Graphs} 
%Community detection is one of the fundamental tasks on graphs where individuals are grouped into clusters or communities based on their graph connectivity. 

\textbf{Community Detection on Temporal Graph.}
In this work, we aim to examine changes to the community structure of the user-user mentions network around the time of COP events. To this end, we utilize the fast and scalable Louvain algorithm~\cite{blondel2008fast} to detect communities on individual snapshots of the temporal graph. The Louvain algorithm optimizes the \emph{modularity} $Q$ of the graph, defined as follows in a directed graph,
\begin{equation}
    Q = \frac{1}{2m} \sum_{ij} [\mat{A}_{ij} - \frac{k_i k_j}{2m}] \delta(c_i, c_j)
\end{equation}
where $\mat{A}$ is the weighed adjacency matrix of the graph, $k_i$ and $k_j$ are the sum of the weights of edges of nodes $i,j$ respectively, $m$ is the sum of all of the edge weights in the graph, $c_i$ and $c_j$ are the communities of nodes $i,j$ and lastly,  $\delta$ is the kronecker delta function. The complexity of the Louvain algorithm is $O(n \cdot \log n )$ where $n$ is the number of nodes. More details can be found in Appendix~\ref{app:louvain}.

\begin{figure}[t]
\centering
\begin{subfigure}{.35\textwidth}
  \centering
  \includegraphics[width=\linewidth]{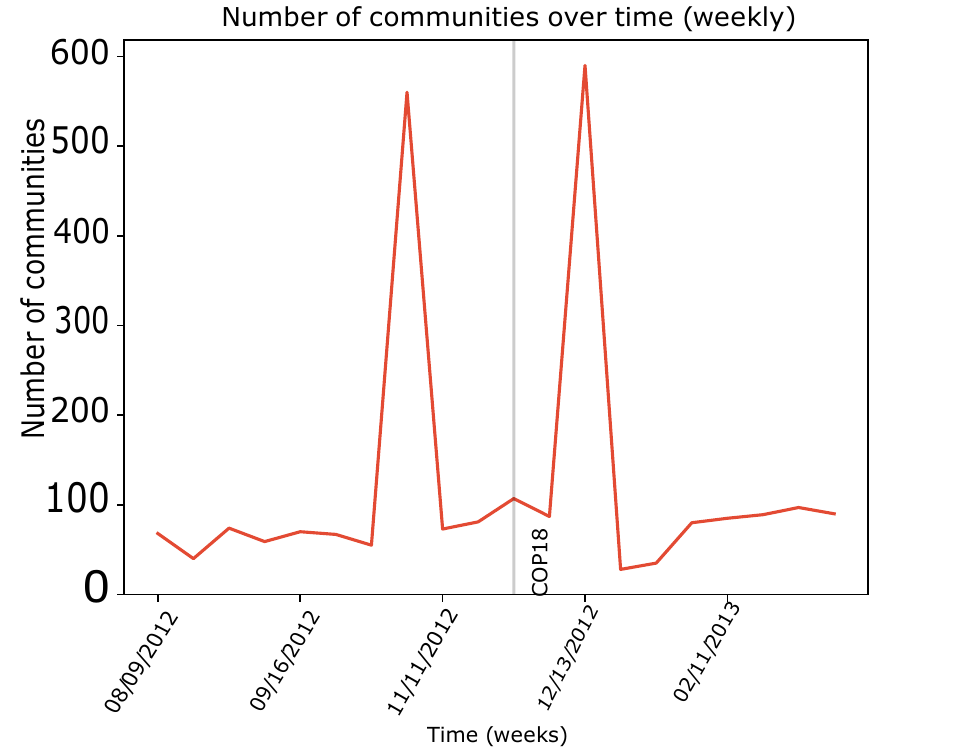}
  \caption{COP 18}
  \label{Fig:cop18}
\end{subfigure}%
\begin{subfigure}{.35\textwidth}
  \centering
  \includegraphics[width=\linewidth]{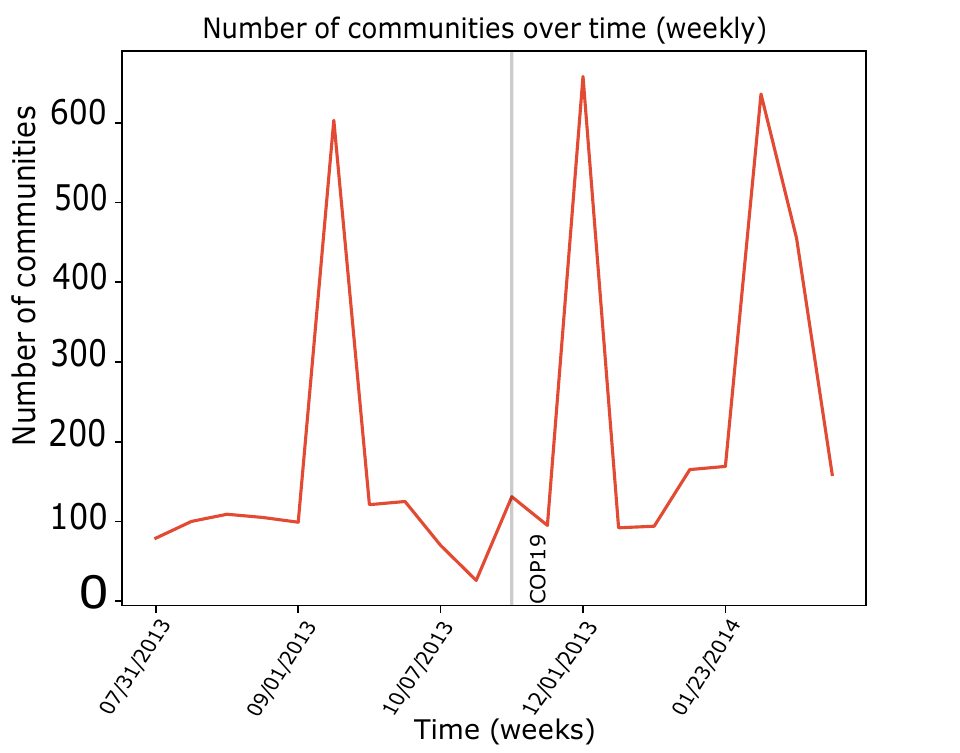}
  \caption{COP 19} 
  \label{Fig:cop19}
\end{subfigure}
\caption{
The number of communities detected around a). COP 18 and b). COP 19} 
\label{Fig:cop18and19}
% \vskip -0.2in
\end{figure}

Overall, The number of communities seem to increase over time with local peaks periodically. We also examine the community structure of graph snapshots surround COP events (10 weeks before and 10 weeks after) with the Louvain algorithm. Figure~\ref{Fig:cop18and19} shows the change in number of communities around COP 18 and COP 19. For COP 18 and 19, there is a decrease in the number of communities. This shows that larger communities are formed likely due to increased user discussions. However, for COP 20 and COP 21, there is an increase of communities. Therefore, it is difficult to clearly draw conclusions based on the number of communities around COP events. This is also expected as the Louvain algorithm is computed based the graph structure alone and doesn't utilize the tweet text and other edge features. 
More community detection results are reported in Appendix~\ref{app:coms}.

% Topic modeling is often used in text analysis applications, such as market research, and social media analytics. 

\textbf{Topic Modeling.}
Topic models are a class of unsupervised machine learning that allows us to extract topics from a large corpus of text data. They are powerful techniques that are commonly used for identifying patterns and themes in large text datasets. A typical topic model works by identifying groups of words that tend to co-occur in the same documents or tweets. These groups of words are known as "topics" and can provide insights into the underlying themes and issues that are present in the data. In our analysis of  opinions on Twitter, we used topic modeling to identify the main topics of conversation related to climate change. By using the mini-batch non-negative matrix factorization (NMF) algorithm \cite{10.1162/NECO_a_00168}, we were able to efficiently extract topics from a large-scale dataset and gain insights into the key themes and issues that are salient to the public.

\begin{figure}[t]
\centering
\includegraphics[width=0.8\linewidth]{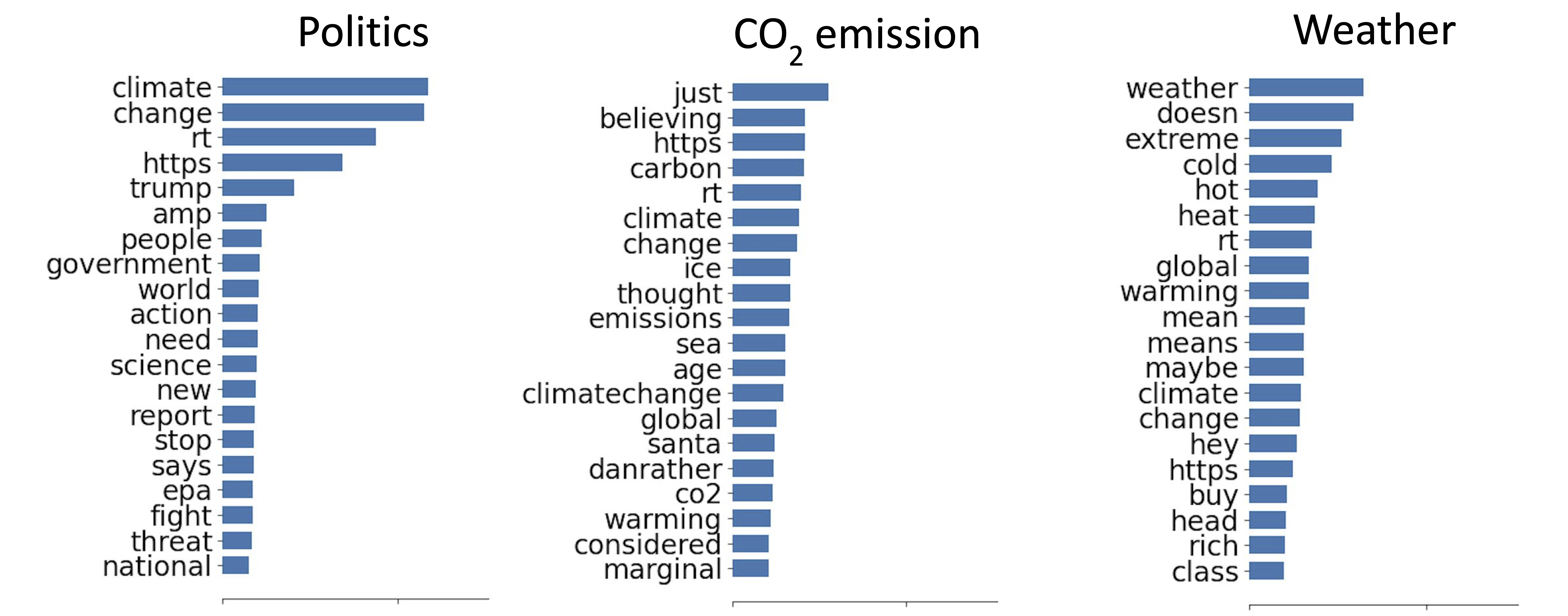}
\caption{Significant topics extracted from the corpus.}
\vskip -0.2in
\label{Fig:topics}
\end{figure}

We first tokenized the tweets data and then trained the mini-batch NMF model on the preprocessed corpus by setting the number of topics as 10. We utilized the scikit-learn toolkit to run mini-batch NMF model. The training algorithm was run for iterations until convergence, i.e., a set of stable topics was obtained.  After the matrix factorization had been done, we looked at the topic-word score matrix $W$ and described each topic by the words with the highest relevance score. 
Figure~\ref{Fig:topics} shows three obtained representative topics with their top 20 words about \emph{politics}, \emph{\ch{CO2} emission}, and \emph{weather}. The \emph{politics} topic is related to politicians (e.g., Trump), organizations (e.g., EPA) and actions to address and mitigate climate change. The \emph{\ch{CO2} emission} topic includes global warming, Arctic ice melting, and sea level rising as consequences of \ch{CO2} emission. The \emph{weather} topic shows people's concern about extreme and abnormal weather caused by global warming.

\section{Pathway to Impact}

 This project aims to provide valuable insights into opinions about climate change on Twitter, potentially impacting many stakeholders. For \emph{policy makers}, our approach provides an understanding of public opinion and sentiment as guidance to decision-making. For \emph{Researchers in social sciences}, our project contributes to the study of the patterns and dynamics of climate change-related online communication. For \emph{journalists}, our study provides insights about when and how to report on climate change topics. For \emph{the public}, this project can inform them about potential negative sentiment and opinions towards climate change events on social media thus help more individuals raise awareness of climate change topics. Future work can extend our work and apply more recent techniques such as Large Language Models~\cite{llm_doi:2303.07205} and Graph Neural Networks~\cite{xu2018powerful}.

\section{Acknowledgement}

We thank Prof. David Rolnick and TA Michelle Lin for their support and guidance throughout the duration of this project. We also thank prof. Reihaneh Rabbany and prof. Guillaume Rabusseau for their discussions and support in this project. This research was supported by the Canadian Institute for Advanced Research (CIFAR AI chair program), Natural Sciences and Engineering Research Council of Canada (NSERC) Postgraduate Scholarship-Doctoral (PGS D) Award and Fonds de recherche du Québec – Nature et Technologies (FRQNT) Doctoral Award.

%We appreciate their dedication to providing constructive feedback and offering precious suggestions that helped us improve the quality of the work.  

% \input{040Experiments}

% \input{030Machine_Learning_Methods}
% \input{040Preliminary_Work}
% \input{050Existing_vs_ML}

%\input{060_NextSteps}

% \input{070General_Considerations}
% \input{99acknowledgement}
% \input{030data}
% \input{040method}
%\input{050pathway}

\bibliography{ref}
\bibliographystyle{abbrv}

\appendix

\newpage
\section{General Considerations}
 We encountered several general considerations that we believe are worth noting.

\textbf{Ethical Considerations:}
Respecting the privacy and confidentiality of Twitter users is crucial, and we have taken steps to de-identify individual users in our analysis and focus on trends of the population instead of investigating individuals. However, we acknowledge that social media data analysis can raise ethical concerns, and urge researchers and practitioners to be mindful of potential harms and take appropriate measures.

\textbf{Data Quality Considerations:}
Slang, buzzwords, and sarcasm can be challenging to interpret in social media data analysis. To address this, we employed natural language processing and sentiment analysis techniques, but acknowledge that our analysis may still be subject to noise and bias.

\textbf{Limitations of Twitter Data:}
While Twitter can provide valuable insights into public opinions, it is important to recognize its limitations. Twitter users may not be representative of the broader population, and the topics and issues discussed on Twitter may not be indicative of those that are most relevant to the public.

\textbf{Opportunities for Future Research:}
Our analysis suggests several avenues for future research on climate-change-related opinion on social media, such as exploring demographic differences in how climate change is perceived and discussed on Twitter, or tracking shifts in public opinion over time.

% \section{Additional Related Work} \label{app:related}

% We discuss additional related work on misinformation detection and political polarization detection in this section.

\section{Louvain Community Detection Algorithm} \label{app:louvain}

The Louvain algorithm has two phases which are interactively applied. First, all nodes in the graph is assigned its own unique community. Then, the algorithm decides if moving a node from its own community to its neighbors' community improves the modularity score $Q$. The node is then placed into its neighbor's community which sees the most increase in modularity. In the second phase, the algorithm groups all nodes in the same community and builds a new network where nodes are the communities from before. After this creation, the first phase then starts again. The algorithm terminates when no modularity increase can be observed. The complexity of the Louvain algorithm is $O(n \cdot \log n )$ where $n$ is the number of nodes. Note that by design, the Louvain algorithm constructs hierarchical communities as it iteractively groups communities into larger ones.

\section{COP Events Details} \label{app:cop_events}

\begin{table}[t] 
\caption{Conference of the Parties~(COP) Events present in the Twitter dataset.}
    { 
    \begin{center}
    \begin{tabular}{| l | l | l |}
        \toprule
         Name & Date & Location \\
        \midrule
        COP13 & 03 Dec 2007 - 17 Dec 2007 & "Bali, Indonesia" \\
        COP14 & 01 Dec 2008 - 12 Dec 2008 & "Poznan, Poland" \\
        COP15 & 07 Dec 2009 - 18 Dec 2009 & "Copenhagen, Denmark" \\
        COP16 & 28 Nov 2010 - 10 Dec 2010 & "Cancun, Mexico" \\
        COP17 & 28 Nov 2011 - 09 Dec 2011 & "Durban, South Africa" \\
        COP18 & 26 Nov 2012 - 07 Dec 2012 & "Doha, Qatar" \\
        COP19 & 11 Nov 2013 - 23 Nov 2013 & "Warsaw, Poland" \\
        COP20 & 01 Dec 2014 - 12 Dec 2014 & "Lima, Peru" \\
        COP21 & 30 Nov 2015 - 12 Dec 2015 & "Paris, France" \\
        COP22 & 07 Nov 2016 - 18 Nov 2016 & "Marrakech, Morocco" \\
        COP23 & 06 Nov 2017 - 17 Nov 2017 & "Bonn, Germany" \\
        COP24 & 03 Dec 2018 - 14 Dec 2018 & "Katowice, Poland" \\
        COP25 & 02 Dec 2019 - 13 Dec 2019 & "Madrid, Spain" \\
        \bottomrule
    \end{tabular}
    \end{center}
    \label{tab:coplist}
    }
\end{table}
We report the COP events present in the Twitter dataset in Table~\ref{tab:coplist}.

\section{Additional Community Detection Results} \label{app:coms}

\begin{figure}[t]
\centering
\includegraphics[width=0.8\linewidth]{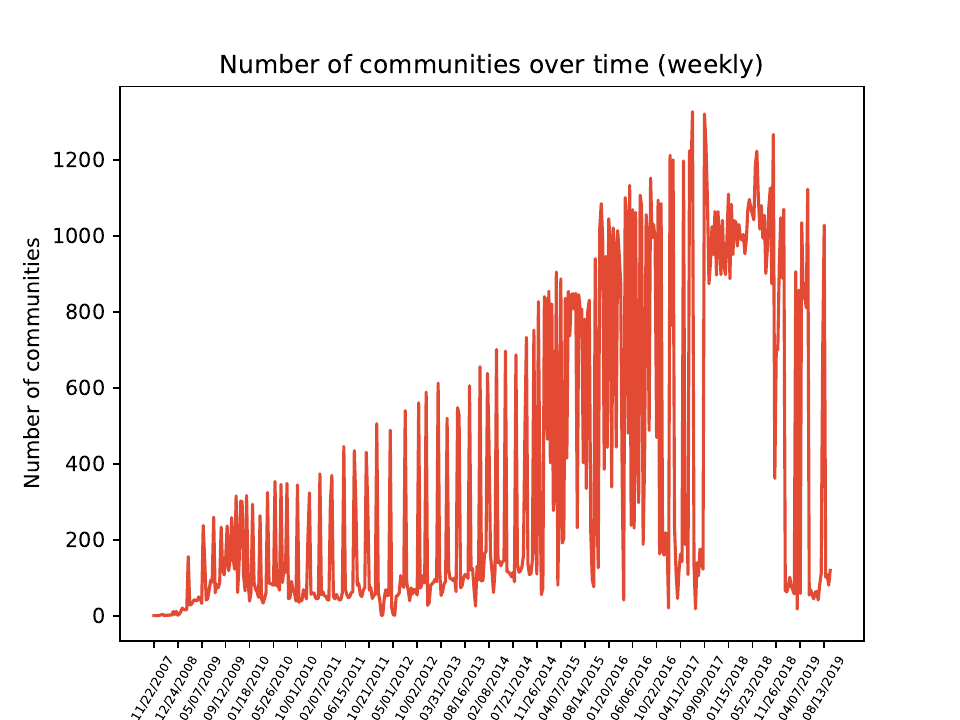}
\caption{The evolution of number of clusters in each snapshot of the temporal graph.}
\vskip -0.2in
\label{Fig:cluster}
\end{figure}

\begin{figure}[t]
\centering
\begin{subfigure}{.35\textwidth}
  \centering
  \includegraphics[width=\linewidth]{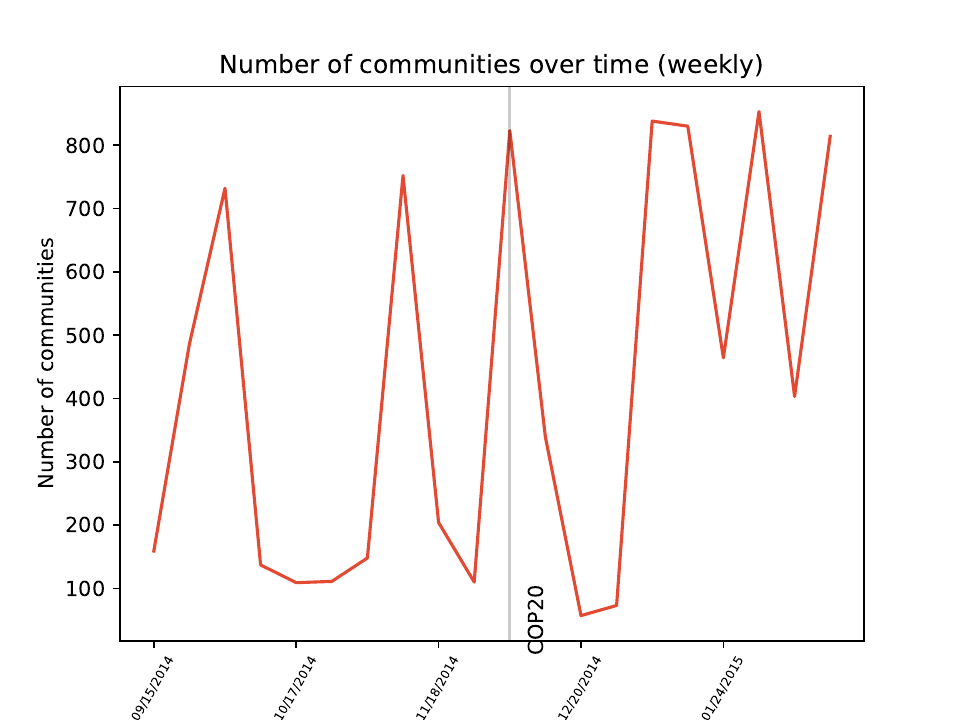}
  \caption{COP 20}
  \label{Fig:cop20}
\end{subfigure}%
\begin{subfigure}{.35\textwidth}
  \centering
  \includegraphics[width=\linewidth]{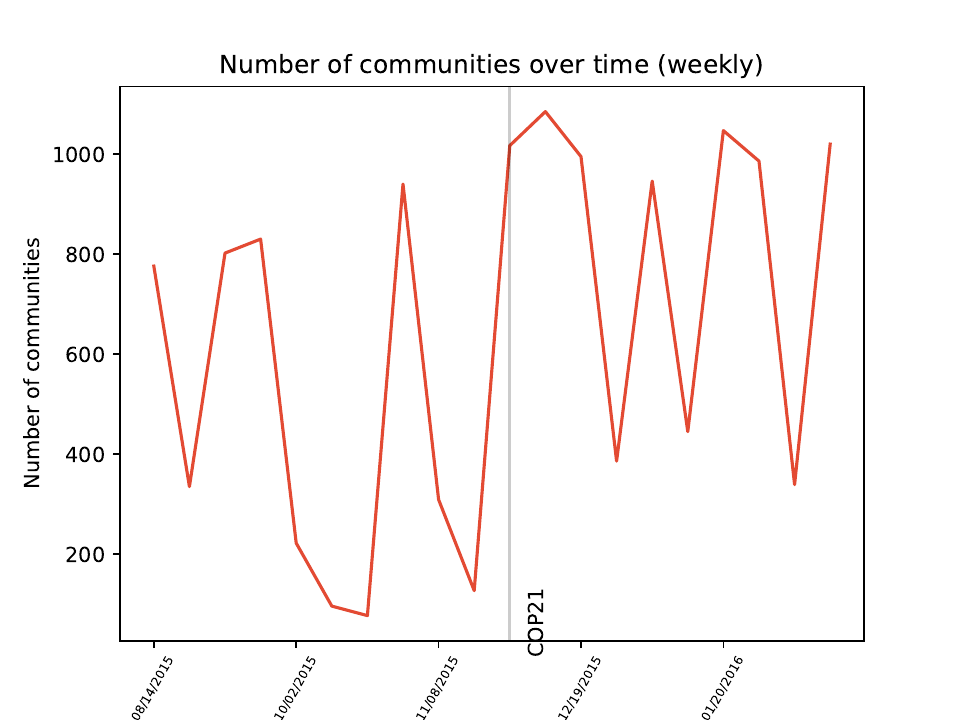}
  \caption{COP 21} 
  \label{Fig:cop21}
\end{subfigure}
\caption{
The number of communities detected by the Louvain algorithm around a). COP 20 and b). COP 21 event.} 
\label{Fig:cop20and21}
\vskip -0.2in
\end{figure}

\begin{figure}[t]
\centering
\begin{subfigure}{.4\textwidth}
  \centering
  \includegraphics[width=\linewidth]{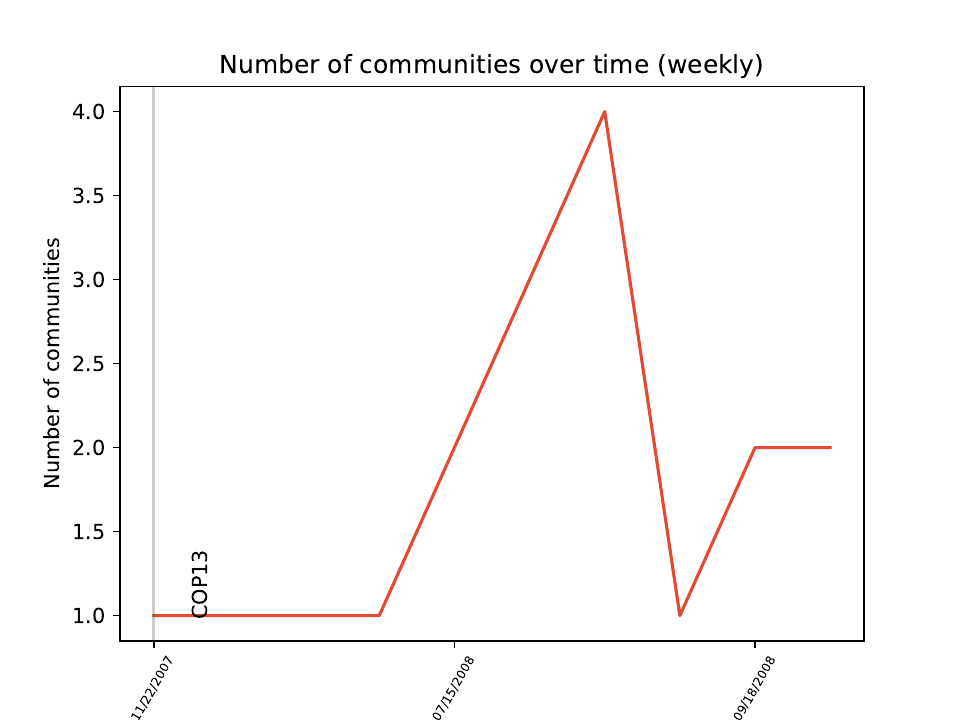}
  \caption{COP 13}
\end{subfigure}%
\begin{subfigure}{.4\textwidth}
  \centering
  \includegraphics[width=\linewidth]{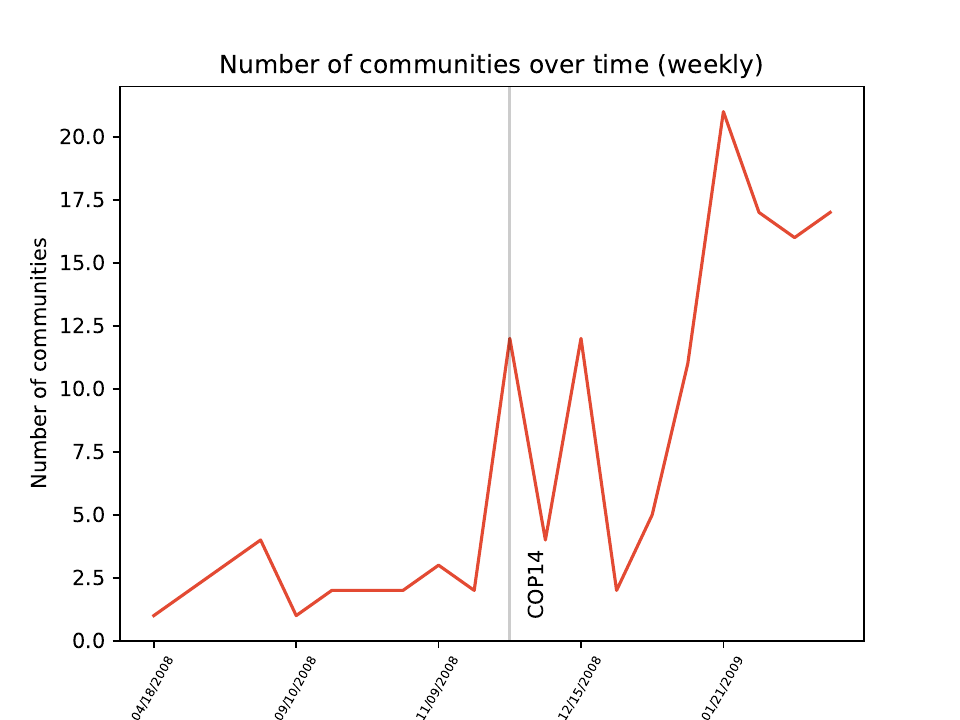}
  \caption{COP 14} 
\end{subfigure}
\begin{subfigure}{.4\textwidth}
  \centering
  \includegraphics[width=\linewidth]{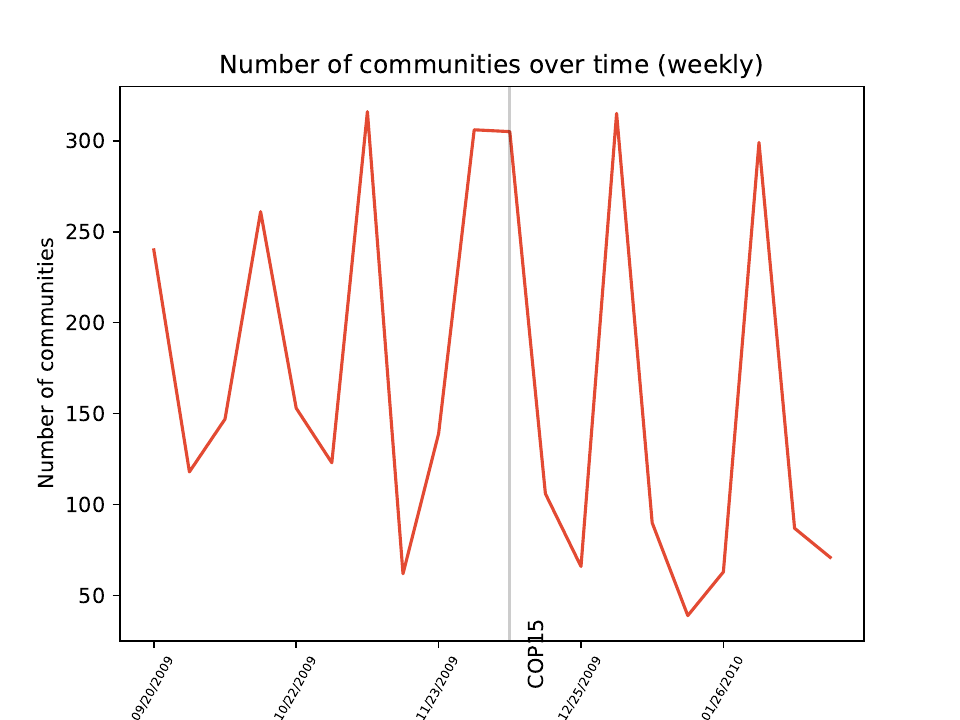}
  \caption{COP 15}
\end{subfigure}%
\begin{subfigure}{.4\textwidth}
  \centering
  \includegraphics[width=\linewidth]{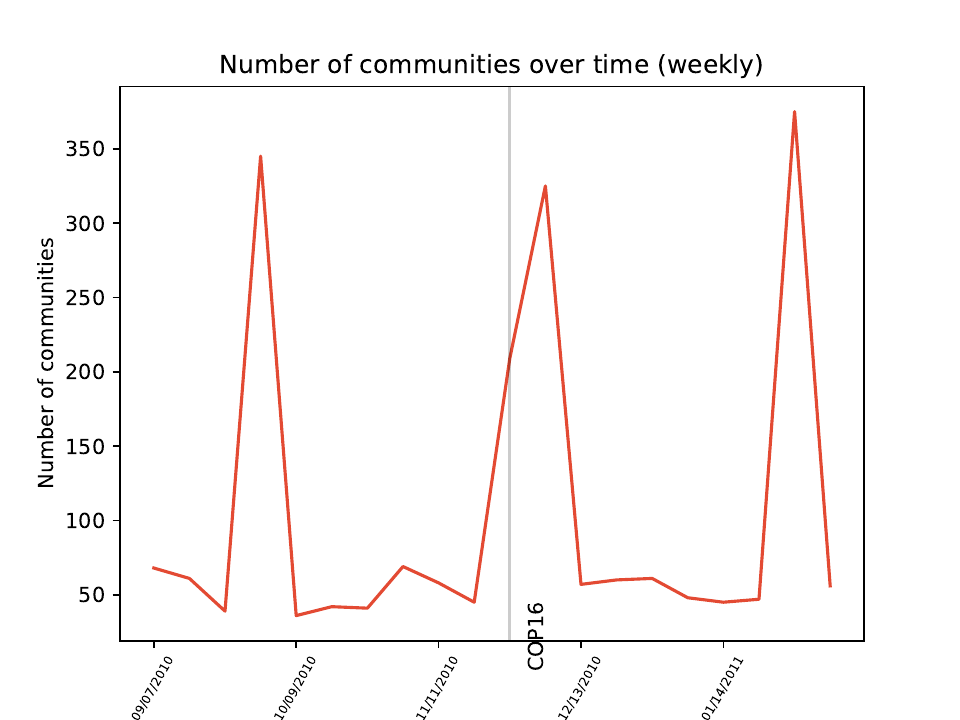}
  \caption{COP 16} 
\end{subfigure}
\begin{subfigure}{.4\textwidth}
  \centering
  \includegraphics[width=\linewidth]{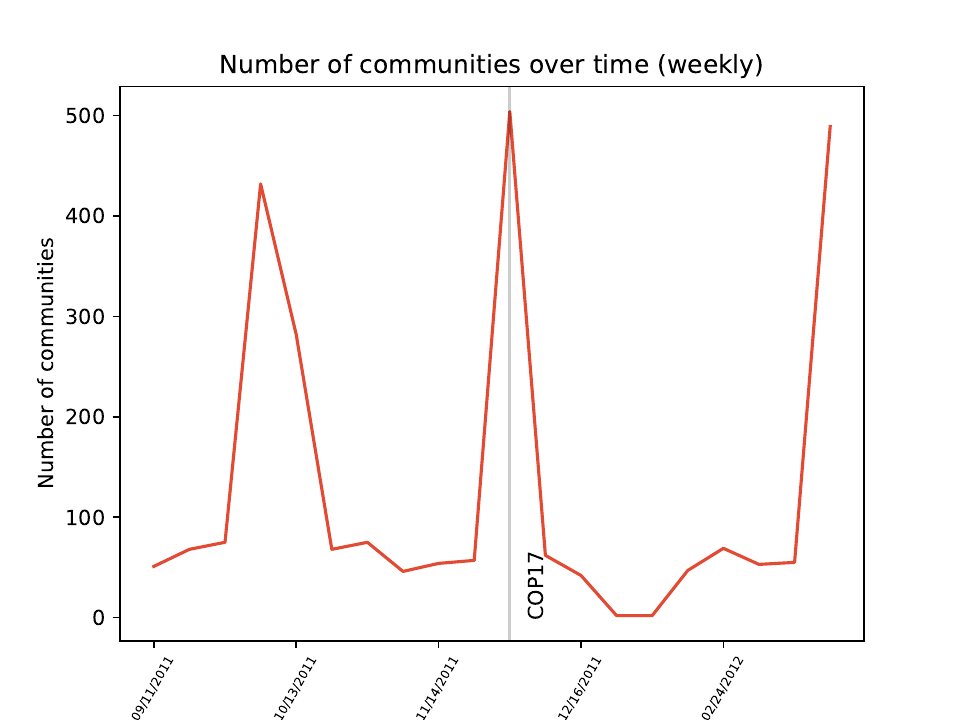}
  \caption{COP 17}
\end{subfigure}%
\begin{subfigure}{.4\textwidth}
  \centering
  \includegraphics[width=\linewidth]{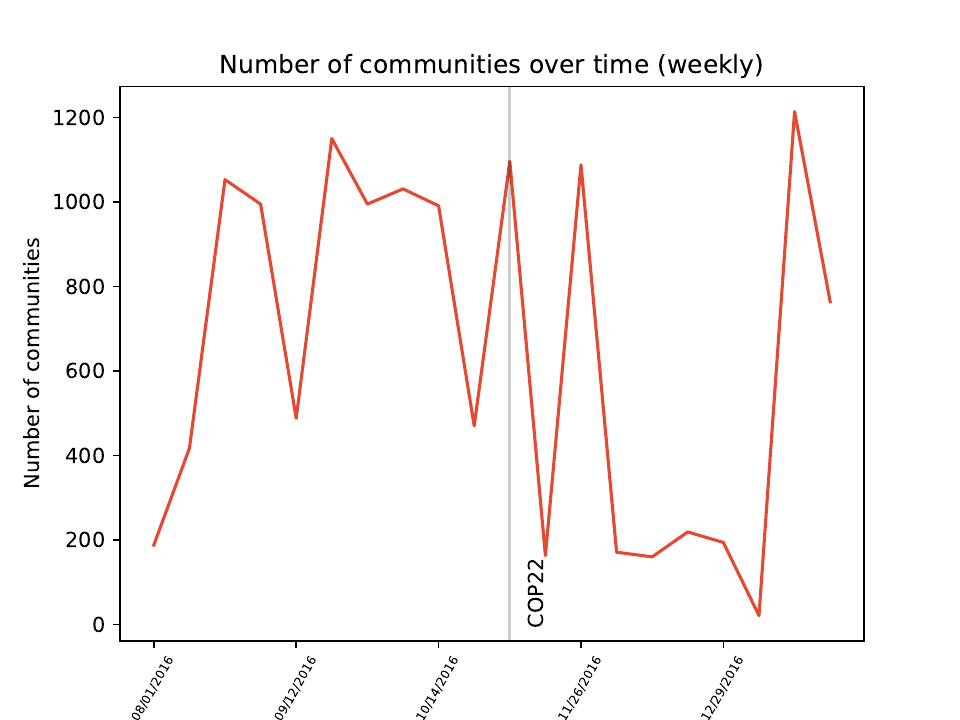}
  \caption{COP 22} 
\end{subfigure}
\begin{subfigure}{.4\textwidth}
  \centering
  \includegraphics[width=\linewidth]{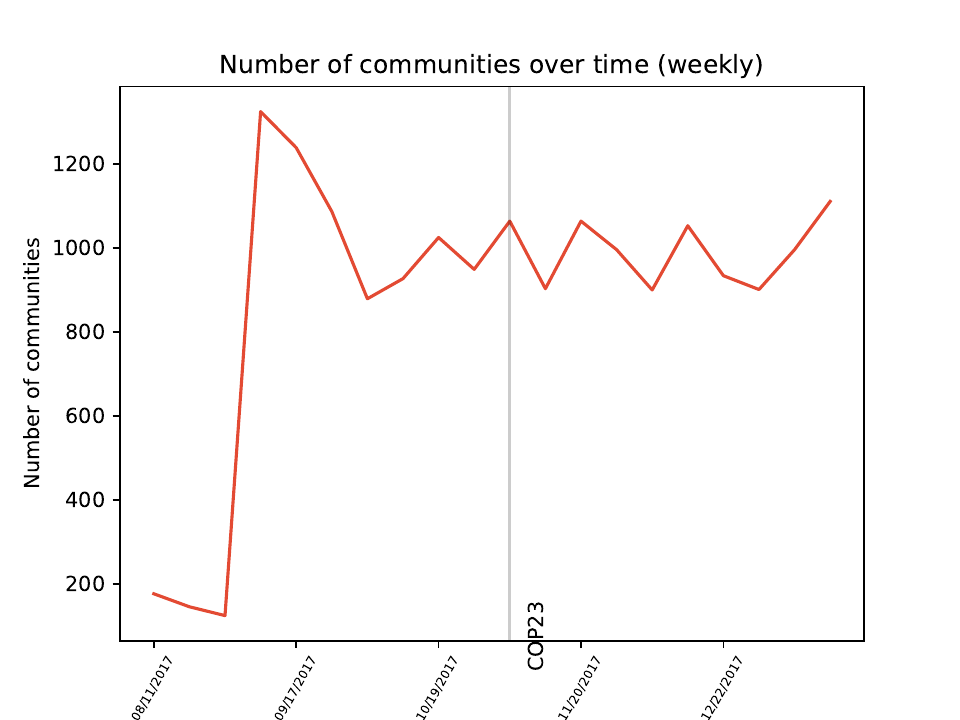}
  \caption{COP 23}
\end{subfigure}%
\begin{subfigure}{.4\textwidth}
  \centering
  \includegraphics[width=\linewidth]{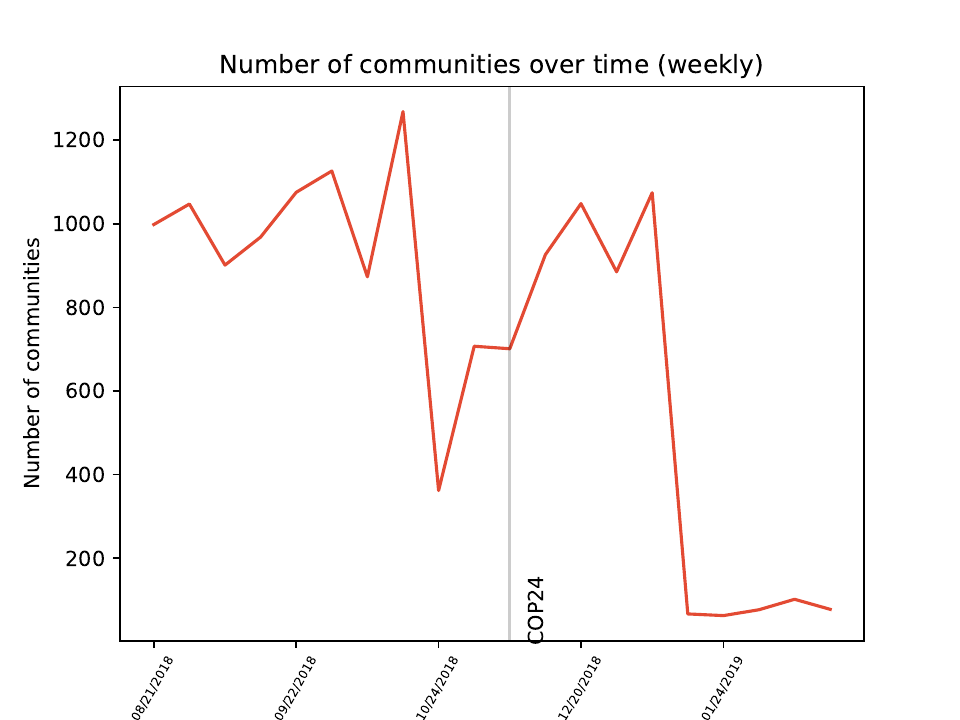}
  \caption{COP 24} 
\end{subfigure}
\caption{
The number of communities detected by the Louvain algorithm around} 
\label{Fig:allcops}
\vskip -0.2in
\end{figure}

Figure~\ref{Fig:cluster} shows the change in number of communities detected by the Louvain algorithm in the temporal graph. Here we provide additional community mining results for all the COP events present in our dataset including COP 13 to COP 24. 

% We analyzed COP 18 to 21 in Section~\ref{sub:coms}. As seen from Figure~\ref{Fig:allcops}, there is no clear association of the change in number of communities with the COP events. 

\end{document}